 \documentclass[preprint2]{aastex}

\slugcomment{Not to appear in Nonlearned J., 45.}

\shorttitle{The 2014 KIDA network}
\shortauthors{Wakelam et al.}

\begin{document}


\title{The 2014 KIDA network for interstellar chemistry}


\author{V. Wakelam\altaffilmark{1,2}, J.-C. Loison\altaffilmark{3,4}, E. Herbst\altaffilmark{5},  B. Pavone\altaffilmark{1,2},
  A. Bergeat\altaffilmark{3,4}, K. B\'eroff\altaffilmark{6},  M. Chabot\altaffilmark{7},
   A. Faure\altaffilmark{8,9},  D. Galli\altaffilmark{10},  W. D. Geppert\altaffilmark{11}, 
   D. Gerlich\altaffilmark{12}, 
   P. Gratier\altaffilmark{1,2},
   N. Harada\altaffilmark{13},
      K. M. Hickson\altaffilmark{3,4}, 
     P. Honvault\altaffilmark{14,15}, 
   S. J. Klippenstein\altaffilmark{16}, 
   S. D. Le Picard\altaffilmark{17}, 
   G. Nyman\altaffilmark{18}, 
   M. Ruaud\altaffilmark{1,2}, 
 S. Schlemmer\altaffilmark{19}, 
 I. R. Sims\altaffilmark{17}, 
 D. Talbi\altaffilmark{20}, 
 J. Tennyson\altaffilmark{21},
 R. Wester\altaffilmark{22} }

\altaffiltext{1}{Univ. Bordeaux, LAB, UMR 5804, F-33270, Floirac, France}
\altaffiltext{2}{CNRS, LAB, UMR 5804, F-33270, Floirac, France }
\altaffiltext{3}{Univ. Bordeaux, ISM, CNRS UMR 5255, F-33400 Talence, France}
\altaffiltext{4}{CNRS, ISM, CNRS UMR 5255, F-33400 Talence, France}
\altaffiltext{5}{Departments of Chemistry, Astronomy, and Physics,  University of Virginia, Charlottesville, VA 22904 USA}
\altaffiltext{6}{Institut des Sciences Mol\'eculaires d'Orsay, CNRS and Universit\'e Paris-Sud, 91405 Orsay cedex, France}
\altaffiltext{7}{Institut de Physique Nucl\'eaire d'Orsay, IN2P3-CNRS and Universit\'e Paris-Sud, 91406 Orsay cedex, France}
\altaffiltext{8}{Universit\'e de Grenoble Alpes, IPAG, F-38000 Grenoble, France}
\altaffiltext{9}{CNRS, IPAG, F-38000 Grenoble, France}
\altaffiltext{10}{INAF-Osservatorio Astrofisico di Arcetri, Largo E. Fermi 5, I- 50125, Firenze, Italy}
\altaffiltext{11}{Department of Physics, University of Stockholm,Roslagstullbacken 21, S-10691 Stockholm}
\altaffiltext{12}{Department of Physics, Technische Universitaet Chemnitz, Germany, Germany}
\altaffiltext{13}{Academia Sinica Institute of Astronomy and Astrophysics, 11F of Astronomy-Mathematics Building, AS/NTU. No.1, Sec. 4, Roosevelt Rd, Taipei 10617, Taiwan, R.O.C.}
\altaffiltext{14}{Laboratoire Interdisciplinaire Carnot de Bourgogne, UMR CNRS 5209, Universit\'e de Bourgogne, 21078 Dijon Cedex, France} 
\altaffiltext{15}{UFR Sciences et Techniques, Universit\'e de Franche-Comt\'e, 25030 Besan\c{c}on Cedex, France}
\altaffiltext{16}{Chemical Sciences and Engineering Division, Argonne National Laboratory, Argonne, IL, 60439, USA}
\altaffiltext{17}{Institut de Physique de Rennes, UMR CNRS 6251, Universit\'e de Rennes 1, 263 Avenue du G\'en\'eral Leclerc, 35042 Rennes Cedex, France}
\altaffiltext{18}{ Department of Chemistry and molecular biology, University of Gothenburg, 41296 Gothenburg, Sweden}
\altaffiltext{19}{ I. Physikalisches Institut, University of Cologne, Z{\"u}lpicher Str. 77, 50937 Cologne, Germany}
\altaffiltext{20}{Laboratoire Univers et Particules de Montpellier UMR 5299, CNRS et Universit\'e de Montpellier, place Eug\`ene Bataillon, 34095 Montpellier, France}
\altaffiltext{21}{Department of Physics and Astronomy, University College London, Gower St., London WC1E 6BT, UK}
\altaffiltext{22}{Institut f{\"u}r Ionenphysik und Angewandte Physik, Universit{\"a}t Innsbruck, Technikerstra{\ss}e 25, A-6020 Innsbruck, Austria}


\begin{abstract}
Chemical models used to study the chemical composition of the gas and the ices in the interstellar medium are based on a network of chemical reactions and associated rate coefficients. These reactions and rate coefficients are partially compiled from data in the literature, when available. We present in this paper kida.uva.2014,  a new updated version of the kida.uva public gas-phase network first released in 2012. In addition to a description of the many specific updates, we illustrate changes in the predicted abundances of molecules for cold dense cloud conditions as compared with the results of the previous version of our network,  kida.uva.2011.

\end{abstract}


\section{Introduction}

Astrochemical models have been developed over the years to study the chemical composition of the interstellar medium (ISM) \citep[see][and references therein]{2013ChRv..113.8710A}. Current models follow the abundance of hundreds of species through thousands of reactions 
\citep{2013A&A...550A..36M}. Although the processes at the surface of the grains, such as sticking, surface reactions, and desorption, are crucial \citep{2013ChRv..113.8707H}, gas-phase reactions are of prime importance and poor estimates of rate coefficients can induce large errors in the predicted abundances, as occurs for OCS \citep{Loison2012}. The rate coefficients for chemical reactions under the ISM conditions of low density and  temperatures from about 10 K to hundreds of K can be studied in the laboratory for some systems or studied theoretically for others \citep{2011ARA&A..49...29S,2012RPPh...75f6901L}. Because each study on a specific reaction  can take from a few months to several years, such studies are best undertaken for systems identified as the most important ones \citep{2010SSRv..156...13W}.  For other systems, general rules can be used (with caution) to estimate the rate coefficients by similarities to other studied reactions \citep{Loison2014c}.  Even for systems studied in detail, it is uncommon to explore temperatures as low as 10 K, so that some estimate or extrapolation of the temperature dependence may be necessary but must then be performed with caution.

The KIDA online database\footnote{http://kida.obs.u-bordeaux1.fr/} is a compilation of rate coefficients coming from various sources including published papers, and other databases or data sheets, which are of interest for the ISM or the modeling of planetary atmospheres.  For ``important" reactions, the appropriate KIDA experts can be asked to provide data sheets with recommendations on the rate coefficients to be used\footnote{http://kida.obs.u-bordeaux1.fr/datasheets}. KIDA contains about 70 such data sheets and the number is increasing with time. The list of experts in physical chemistry is provided in Table~\ref{experts} with their current coordinates. To help astrochemical users, we also provide a subset of chemical reactions, extracted from KIDA, to be used to model the chemistry in the ISM. The first release of this network, kida.uva.2011,  was reported in \citet{2012ApJS..199...21W} (hereinafter W12).  We are now presenting in this paper an updated version, labelled kida.uva.2014. The chemical updates of the network are described in Section 2 while the impact of the new chemistry is discussed in Section 3. 

\section{Chemical updates}\label{updates}

This new network is based on the previously published kida.uva.2011 network (W12) with the updates posted in the KIDA database between October 2011 and October 2014. The main updates of the network come from:
\begin{itemize}
\item [-] Reactions from the high temperature network of \citet{Harada2010},
\item [-] Photo rates from \citet{1988ASSL..146...49V,Roberge1991,2006FaDi..133..231V,2008CP....343..292V},
\item [-] Rate coefficients for the OCS chemistry from \citet{Loison2012},
\item [-] New data sheets \citep[][and many others]{2013arXiv1310.4350W},
\item [-] A review of the HCN/HNC chemistry by \citet{Loison2014b},
\item [-] A review of the carbon chemistry by \citet{Loison2014c},
\item [-] Branching ratios for reactions forming C$_{n=2-10}^{(0,+)}$, C$_{n = 2- 4}$H$^{(0, +)}$, and C$_3$H$_2^{(0,+)}$ from \citet{Chabot2013}.  
\end{itemize}
In addition to these updates, a number of individual reactions have been added or updated from a list of studies given in the Appendix. New species have been added to the network, and are listed in Table~\ref{new_spec}. A total of 446 rate coefficients have been changed and 1038 new reactions have been added. The final network contains 489 species composed of 13 elements and 7509 reactions.   In addition to the gaseous species,  neutral grains, negatively charged grains, grain-surface  electrons and atomic hydrogen are included. The types of reactions and the formulas used to parameterize the rate coefficients are the same as described in W12. The complete network is available at http://kida.obs.u-bordeaux1.fr/uploads/models/kida.uva.2014.zip with a list of species and the bibliographic reference in  bibtex format.  

As for the previous version, kida.uva.2011, we expect the network to be used for the chemical modeling of the interstellar medium at temperatures between 10 and 300~K. Despite the fact that we added reactions from \citet{Harada2010} for higher temperature conditions, including reactions with activation energy, and reverse reactions, we have not done any additional checks on its appropriateness for temperatures larger than 300~K.
For some of the reactions, the parameter $\gamma$ used in the temperature-dependent Arrhenius-Kooij expression for the rate coefficient -  $\rm k(T) = \alpha (T/300)^\beta \exp(-\gamma/T)$ -  is negative. This sign change will produce strongly incorrect results if the calculation of the rate coefficient is extrapolated outside the temperature range of validity. Those reactions are listed in Table~\ref{gamanegative}. Not extrapolating outside of the range of validity can also have consequences. In fact, the temperature dependence of rate coefficients is usually valid over a certain range of temperature, which can be a crucial piece of information. However, in many cases, the information is only partial and could lead to large errors if not carefully considered. This is particularly true for H-atom transfer reactions where quantum mechanical tunnelling results in strong departure from Arrhenius behavior at low temperatures \citep{2014NatCh...6..141T}. As an example, consider the reaction N + H$_2$ $\rightarrow$ H + NH, which has a very large activation energy. This reaction was already present in the previous OSU database \citep{Harada2010} without any indication of the temperature range and, it is likely that common users would extrapolate the rate coefficient down to 10~K. In KIDA, we  included the original rate coefficient from \citet{Davidson1990} with its measured temperature  range of 1950 - 2850~K. The rate coefficient computed at the minimal temperature (1950 K) would then be $4\times 10^{-13}$~cm$^{3}$ s$^{-1}$ (much larger than the 10~K extrapolation of $4\times 10^{-82}$~cm$^{3}$~s$^{-1}$). If only the measured temperature range for N + H$_2$ (1950 - 2850 K ) were included in the network, the reaction rate coefficient for temperatures lower than the lowest measured temperature (1950 K) would be programmed to remain at the measured value for 1950 K.  Likewise, the rate coefficient at temperatures higher than 2850 K would be programmed to remain at the 2850 K value. In dense cloud conditions where both N and H$_2$ are abundant, even the low 1950 K rate coefficient would be large enough to have a significant effect on the chemistry while the rate coefficient of this reaction at 10~K is clearly negligible. For this reason, in kida.uva.2014, we have used the original rate coefficient and reference but we have changed the temperature range to extend to 10~K to allow extrapolation.  
 For reactions between ions and non-polar neutrals with measurements only at 300 K, we have assumed that the rate coefficients are independent of temperature, as in the Langevin approximation, and have used the measured total rate coefficients and product branching fractions over the temperature range 10-300 K.

\section{Modeling results of the new network}

\subsection{Comparison with kida.uva.2011}

\begin{figure*}
\epsscale{1.6}
\plotone{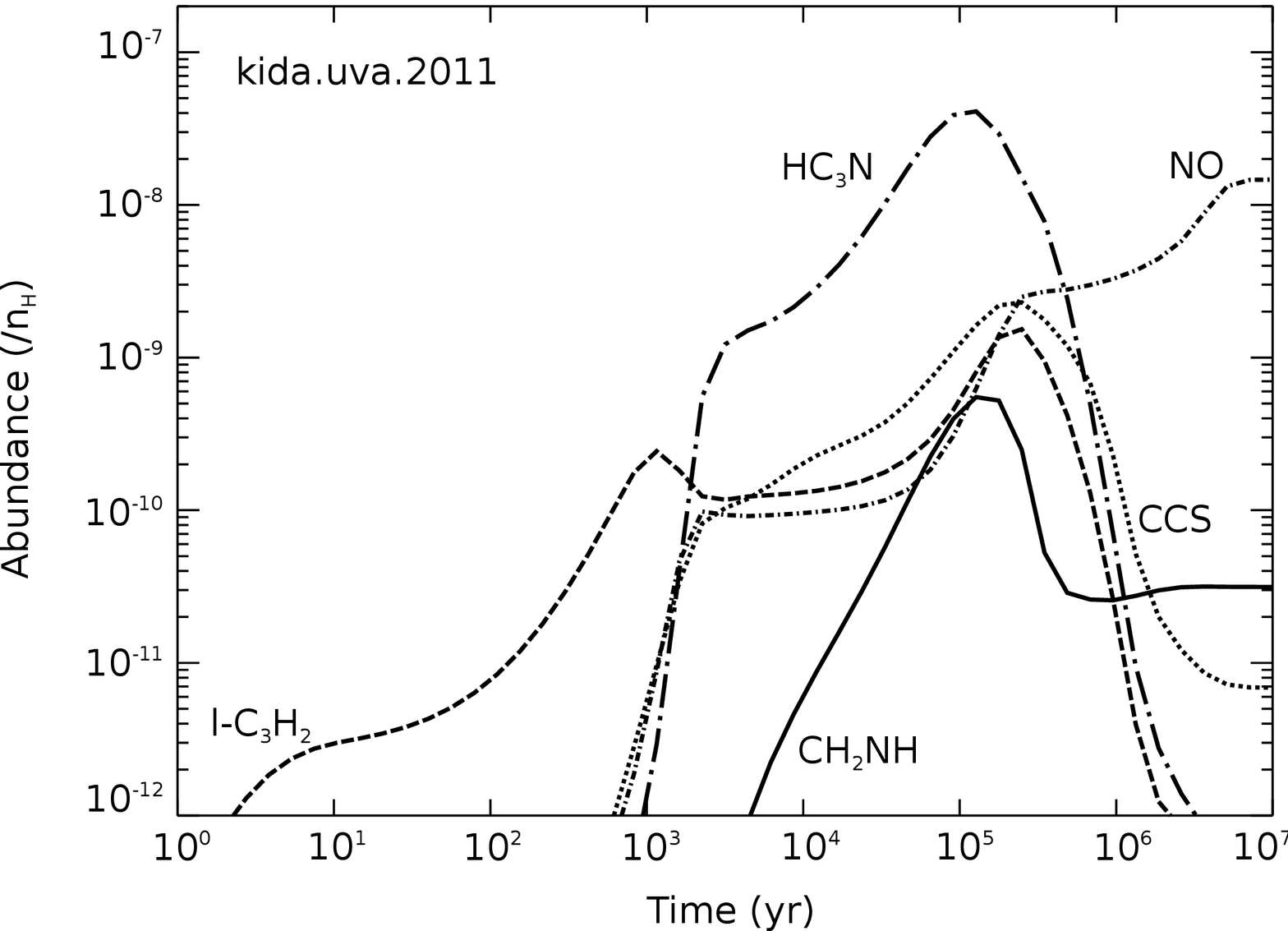}
\plotone{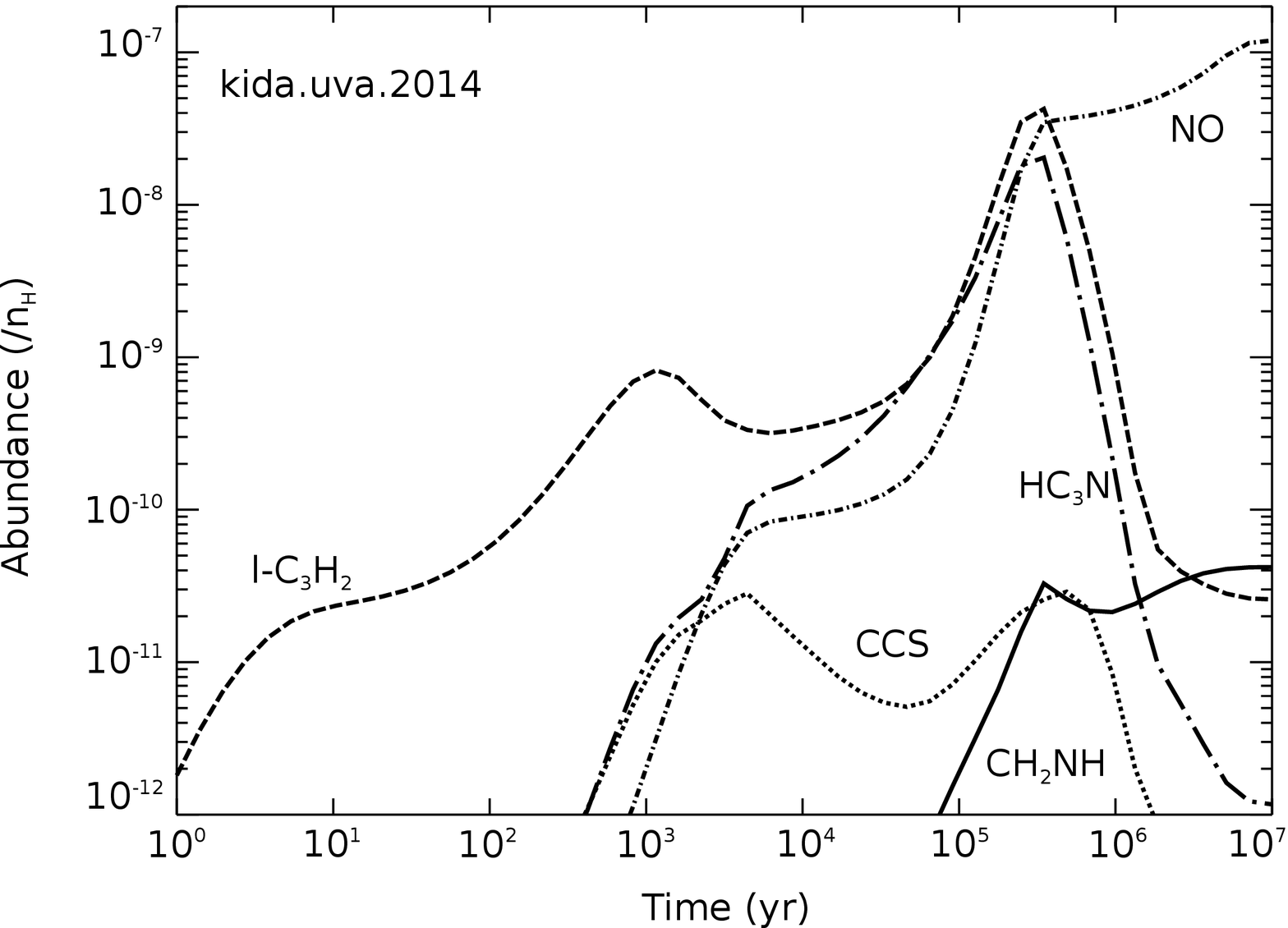}
\caption{Abundances as a function of time for a selection of species for dense cloud conditions using kida.uva.2011 (upper panel) and kida.uva.2014 (lower panel). \label{fig1}}
\end{figure*}

To quantify the impact of the updates done on kida.uva.2011, we have compared the abundances computed with the chemical code Nahoon using kida.uva.2011 (see W12) and kida.uva.2014 with physical parameters typical for dense clouds: a temperature of 10~K, a total ``proton'' density, defined by the relation $n_{\rm H} = n(H) + 2n(H_2)$,  of $2\times 10^4$~cm$^{-3}$, a cosmic-ray ionisation rate for H$_{2}$  of $10^{-17}$~s$^{-1}$ and a visual extinction of 30. The gas-phase abundances of the species are computed as a function of time starting from an atomic composition except for hydrogen, which is initially molecular. The elemental composition utilised is the same as used by \citet{2011A&A...530A..61H}, except for elemental oxygen, which has an abundance with respect to H of  $2.4\times 10^{-4}$. 
Tables~\ref{diff1} and ~\ref{diff2} give lists of abundant species, defined as those with abundances computed with at least one network larger than $10^{-12}$ compared with the total proton density.  The lists  contain only those abundant species with abundances  modified by more than a factor of 2 at $10^5$ and $10^6$~yr with the change in reaction network.  For each of these species, we also give the ratio of the abundance computed with kida.uva.2014 to the one computed with kida.uva.2011 in logarithmic terms. In Fig.~\ref{fig1}, we show the abundances of a selection of species as  functions of time computed with the two networks. 

At $10^5$~yr, the species that show a difference larger than two orders of magnitude are HC$_2$N$^+$, HC$_{\rm n}$N (with n=4, 6 and 8), CH$_2$NH and CCS, and they are all decreased by the new network. The HC$_2$N$^+$ abundance is decreased by the newly added reaction H$_2$ + HC$_2$N$^+$ $\rightarrow$ H + CH$_2$CN$^+$ while the HC$_{\rm n}$N species and CH$_2$NH are decreased by new destruction reactions with atomic carbon, all from \citet{Loison2014b}. The strong decrease of the CCS abundance is due to the introduction of the destruction reaction O + CCS $\rightarrow$ CO + CS, which possesses a rate coefficient of $10^{-10}$~cm$^{3}$~s$^{-1}$ independent of temperature  \citep{Loison2012}. The rate coefficient has been deduced from the similar O + CCO reaction for which the rate coefficient has been measured to be $9\times 10^{-11}$~cm$^3$~s$^{-1}$ \citep{1972JChPh..57.3933S,Bauer1985} at room temperature. The high value of the rate coefficient is compatible with the fact that the O + CCS reaction is a radical-radical reaction which usually shows no (or a submerged) barrier.  Note that this reaction was also included in \citet{Harada2010} but with an activation barrier based on the rate coefficient from O + CS. 

A significant fraction of the molecules listed in Table~\ref{diff1} are either nitrogen-bearing species or carbon-bearing species because a large number of the rate coefficients for these species have been changed, as mentioned in  Section \ref{updates}. The abundances of large carbon-chains have also been reduced. At $10^6$~yr, the list of species is much smaller than at $10^5$~yr and the differences are smaller than at earlier times, partly because the abundance of atomic carbon decreases with increasing time. Among the species that present a variation of more than a factor of 10,  only NO was not already present in Table 3.  This molecule is increased by approximately a factor of 10 because of the decrease in the N + NO rate coefficient as recommended by the KIDA experts in a datasheet.

\subsection{Comparison of  model predictions with observations} 

Although this is not the primary goal of this paper, we have compared the model results, discussed in the previous section, with the observations in TMC-1, as listed in \citet[][Table 4]{2013ChRv..113.8710A}. We initially used only gas-phase chemistry and in a second attempt, we  considered grain-surface chemistry as well. For this second test, we used the Nautilus gas-grain code as described in many previous papers \citep{2010A&A...522A..42S,2014MNRAS.440.3557R,Loison2014c}. The grain surface network was adapted to include the new species introduced in kida.uva.2014 but is basically the same as in \citet{Loison2014c}. It consists mainly of diffusive Langmuir-Hinshelwood processes with adsorption and desorption mechanisms. In addition to thermal desorption, Nautilus  includes cosmic-ray induced desorption, as described in \citet{1993MNRAS.261...83H} and the desorption by exothermicity of surface reactions following \citet{2007A&A...467.1103G} with the parameter $a$ set to 0.01.

\begin{figure*}
\epsscale{1.3}
\plotone{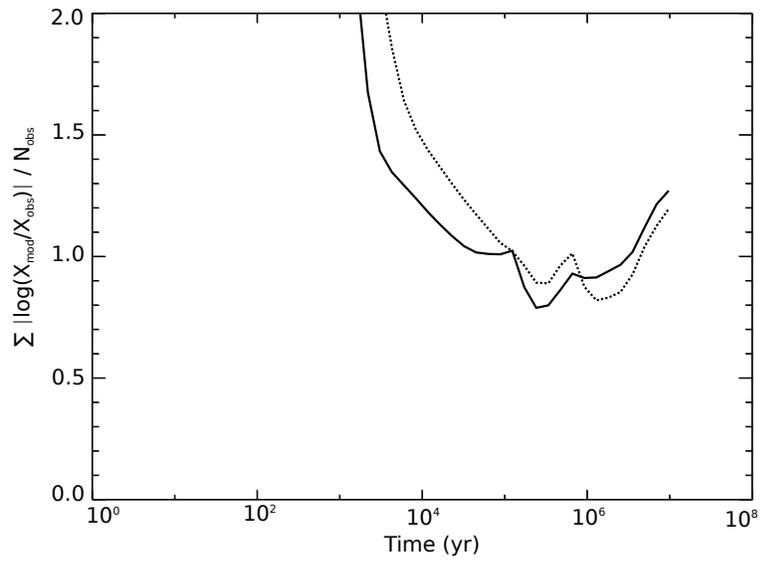}
\caption{ Mean logarithmic difference between the modeled and observed abundances over all observed species (see text) as a function of time using kida.uva.2011 (solid line) and kida.uva.2014 (dashed line). \label{sum_diff_TMC1}}
\end{figure*}

\begin{figure*}
\epsscale{1.6}
\plotone{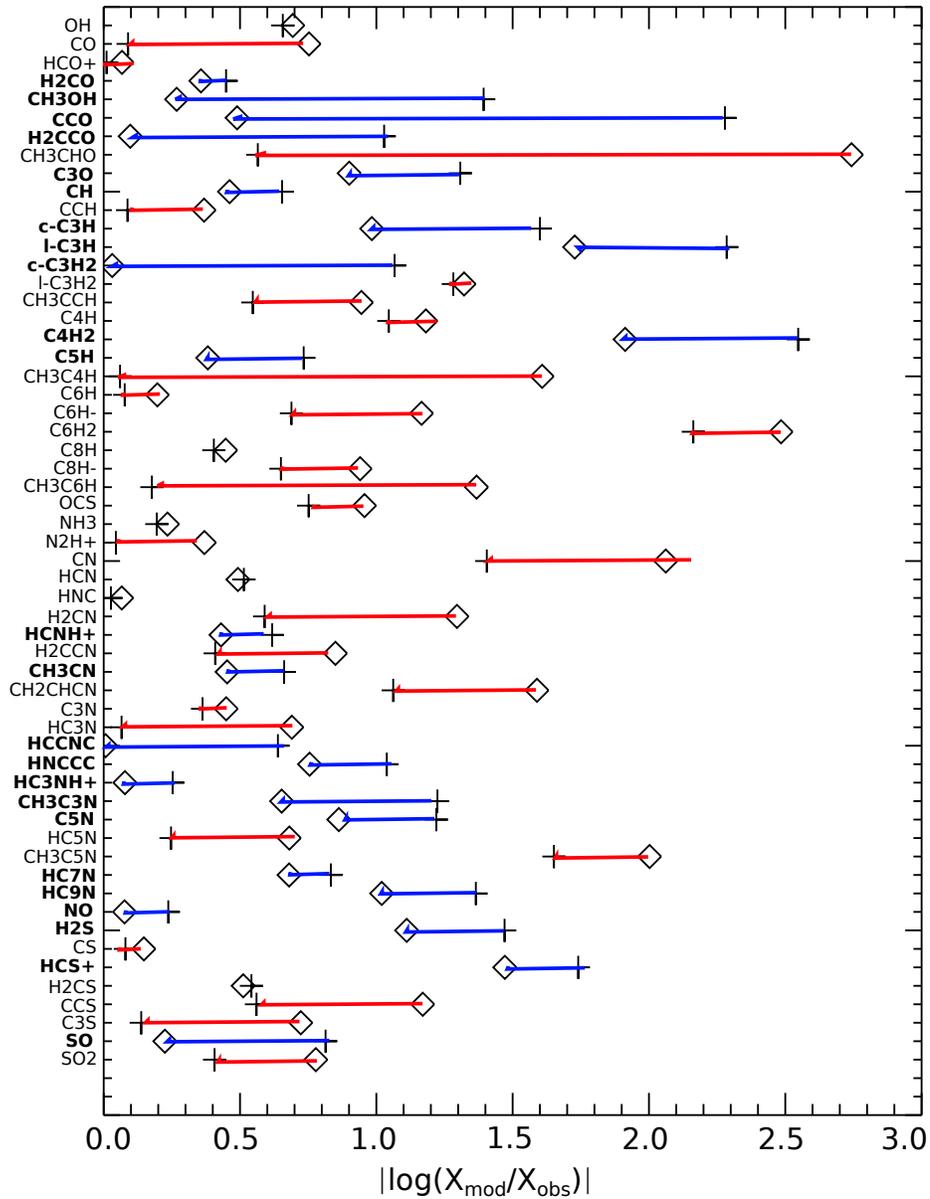}
\caption{ Absolute logarithmic difference between modeled and observed abundances for each species at the best time of each model using plus signs for kida.uva.2011 and diamonds for kida.uva.2014. Horizontal lines show the differences between the old and new values. Red lines indicate that the difference has been increased by the new network while blue lines indicate that it has been reduced. Boldface species are those with abundances improved by the new network. \label{comp_obs_TMC1}}
\end{figure*}

The physical conditions and initial abundances are the same as described in the previous section. To compare with the observations, we have simply computed the  differences between the logarithms of observed and modeled abundances using the two networks,  and determined the time for which the mean logarithmic difference is at its minimum \citep{Loison2014c}.  This logarithmic parameter is defined by the sum over all species of $\left | \log(\rm X_{mod}/ X_{obs})\right | / N_{obs}$,  where $\rm X_{mod}$ and  $\rm X_{obs}$  are the modeled and observed abundances while $\rm N_{obs} $ is the number of observed species (57) in TMC-1 used for the comparison. Considering only gas-phase chemistry, the general agreement with the model is slightly worse with the new network. When surface chemistry is considered, we obtain a similar degree of agreement but for different times. We now only discuss the comparison using the full gas-grain networks. Fig.~\ref{sum_diff_TMC1} shows the mean logarithmic difference as a function of time for both networks.   We have ignored the species for which only upper limits have been constrained. At the best times, for both models, the mean difference between modeled and observed abundances is smaller than a factor of 10. As already noted by \citet{Loison2014c}, we now obtain a reasonable agreement with the observations in TMC-1 using a C/O elemental ratio smaller than 1. The best time is  $2.5\times 10^5$~yr using kida.uva.2011 while it is approximately $1.3\times 10^6$~yr using the new network. For these best times, we have plotted in Fig.~\ref{comp_obs_TMC1} the difference between modeled and observed abundances for individual species. The difference between modeled and observed abundances is decreased for about half of the species and  increased for the other half when comparing the two networks. This explains the fact that the general agreement is not changed. 

\begin{figure*}
\epsscale{1.3}
\plotone{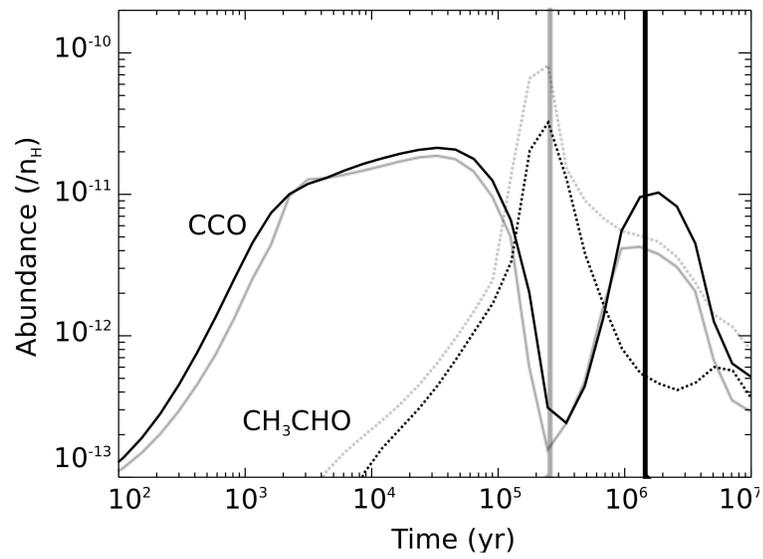}
\caption{Abundances of CH$_3$CHO and CCO as a function of time computed with the kida.uva.2011 (grey lines) and kida.uva.2014 (black lines) networks. The vertical grey and black lines correspond to the best times constrained with the old and new networks respectively. \label{CH3CHO_CCO_2models}}
\end{figure*}

 The two species for which the changes are the strongest are CH$_3$CHO, for which the agreement with the new model is worse,  and CCO, for which agreement with the new model is better. Fig.~\ref{CH3CHO_CCO_2models} shows the abundance of these two species as a function of time predicted with the two networks. The TMC-1 abundance of CCO is $3\times 10^{-11}$ (compared to the total proton density) \citep{1998FaDi..109..205O} while that of CH$_3$CHO is $3\times 10^{-10}$ \citep{1985ApJ...290..609M}. Both modeled abundances are smaller than the observed ones at all times. The CCO predicted abundance is much smaller at $2.5\times 10^5$~yr, the best time using the older network, than at $1.3\times 10^6$~yr. After $10^6$~yr however, the CCO abundance is also larger with the new network because of the larger abundance of atomic carbon  since CCO is produced by C + HCO $\rightarrow$ CCO + H. The better agreement between the new model and the observed CCO abundance is then due to the change in the best time rather than the chemistry itself. For CH$_3$CHO, the worse agreement with the new model is also due to the change in the best time: at later times, the CH$_3$CHO abundance is smaller. The decrease in the gas-phase abundance of this species is however much stronger with the new network as can be seen in Fig.~\ref{CH3CHO_CCO_2models}. At $10^6$~yr, one of the main formation reactions of CH$_3$CHO is O + C$_2$H$_5$ $\rightarrow$ H + CH$_3$CHO. The CH$_3$CHO abundance is then decreased with  a decrease in the C$_2$H$_5$ abundance. The C$_2$H$_5$ molecule is more strongly destroyed using the new network because of the newly introduced reaction H + C$_2$H$_5$ $\rightarrow$ CH$_3$ + CH$_3$ from \citet{Baulch1992}. The rate coefficient for this reaction is $6\times 10^{-11}$~cm$^{-3}$~s$^{-1}$ and was recommended by \citet{Baulch1992} for temperatures between 300 and 2000~K. Here we use the same rate coefficient down to 10~K.

\section{Conclusion}

The KInetic Database for Astrochemistry (KIDA) aims at providing to the community gas-phase kinetic data to study the chemistry under the extreme conditions of the interstellar medium. From this database, a subset of reactions is extracted to be directly used in chemical models. The first release of this subset, kida.uva.2011, was included in W12. After a large number of updates, we are now presenting the new version: kida.uva.2014. A large number of gas-phase species are affected by these modifications at typical cloud chemical ages ($10^5 - 10^6$~yr). The mean ratio between modeled abundances and the ones observed in the TMC-1 dark cloud is  a factor of 6 at best, using a full gas-grain model. This general agreement is not much affected by the new gas-phase network but the "best time" is older than obtained with the previous one. 

\acknowledgments

The maintenance and developments of the KIDA database are possible thanks to the European Research Council Starting Grant 3DICE (grant agreement 336474), the French program PCMI and the Observatoire Aquitain des Sciences de l'Univers. The KIDA team is also grateful to the persons who submitted data to the database. S.D.L.P. is grateful to the Institut Universitaire de France, while E. H. thanks the NSF and NASA for support. The work at Argonne is supported by the U.S. Department of Energy, Office of Science, Office of Basic Energy Sciences, Division of Chemical Sciences, Geosciences, and Biosciences under Contract No. DE-AC02-06CH11357.


\section{Appendix}

\subsection{References for updates}

The references used to update the network are: \citet{Adams1984, Adusei1996, Anicich1993, Arthur1978, Avramenko1967, Azatyan1975, Bauerle1995a, Bauerle1995b, Baulch1992, Baulch1994, Bohland1985, Bryukov2001, Chabot2013, Cohen1991, Cole2012, Daranlot2011, Daranlot2012, 2013PCCP...1513888D, Diau1995, Dombrowsky1992, Frank1986, Freund1977, Fujii1987, Gustafsson2012, Hack1979, Hamberg2010, Hanson1984, Harada2010, Hemsworth1974, Henshaw1987, Herbrechtsmeier1973, Hickson2013, Humpfer1995, Karkach1999, Kim1975a, Kim1975b, Leen1988, Lifshitz1991, Lloyd1974, Loison2012, Loison2014a, Loison2014b, Loison2014c, Mackay1981, Mackay1980, MagnusGustafsson2014, Maluendes1993, Mayer1966, Mebel1996, Millar1986, Millar1987, Millar1991, Miller1988, Mitchell1984a, Mitchell1984b, Miyoshi1993, Murrell1986, Neufeld2009, Nguyen2004, OttoR.2008, Patterson1962, Payzant1975, Peeters1973, Roberge1991, Rodgers1996, Schofield1973, Shannon2014, Singleton1988, Smith1984, Smith1992, Smith1994, Stewart1989, Tsang1992, Tsang1986, Tsang1991, Tsuboi1981, 1988ASSL..146...49V, 2006FaDi..133..231V, 2008CP....343..292V, Vandooren1994, Warnatz1984, Yang1993, Zanchet2009, Zellner1988}.



\clearpage


\begin{deluxetable}{ccc}
\tabletypesize{\small}
\tablecolumns{3}
\tablecaption{List of KIDA experts\label{experts}}
\tablehead{
\colhead{Expert name} & \colhead{Contact information} & \colhead{Field of expertise}
}
 \startdata
  Astrid Bergeat & University of Bordeaux, France & (2) \\
   & astrid.bergeat@u-bordeaux.fr & \\
  Karine B\'eroff & University Paris-Sud, France & (1, 2, 4, 5) \\ 
  & karine.beroff@u-psud.fr & \\
  Marin Chabot & University Paris-Sud, France & (1, 2, 4, 5) \\
  & chabot@ipno.in2p3.fr & \\
 Alexandre Faure & University of Grenoble, France & (2) \\
 & Alexandre.Faure@obs.ujf-grenoble.fr & \\
 Daniele Galli & INAF Arcetri Astrophysical Observatory, Italy & (1)\\
  & galli@arcetri.astro.it & \\
 Wolf D. Geppert & University of Stockholm, Sweden & (2, 5, 6) \\
 & wgeppert@hotmail.com & \\
  Dieter Gerlich & Technische Universitaet Chemnitz, Germany & (2, 3, 4, 6) \\
  & gerlich@physik.tu-chemnitz.de & \\
 Eric Herbst & University of Virginia, USA & (2, 3, 5) \\
 & ericherb@gmail.com & \\
 Kevin M. Hickson & University of Bordeaux, France & (1,2, 5) \\
 & km.hickson@ism.u-bordeaux1.fr & \\
 Pascal Honvault &  University of Franche-Comt\'e / Bourgogne, France & (2) \\
 & pascal.honvault@univ-fcomte.fr & \\
 Stephen J. Klippenstein & Argonne National Laboratory, USA & (2, 3) \\
 & sjk@anl.gov & \\
 S\'ebastien D. Le Picard & University of Rennes, France & (2) \\
 & sebastien.le-picard@univ-rennes1.fr & \\
 Jean-Christophe Loison & University of Bordeaux, France & (1, 2, 3, 5) \\
 & jc.loison@ism.u-bordeaux1.fr & \\
  Gunnar Nyman & University of Gothenburg, Sweden & (2, 3) \\
  & nyman@chem.gu.se & \\
Stephan Schlemmer & University of Cologne, Germany & (2, 3) \\
& schlemmer@ph1.uni-koeln.de &\\
  Ian R. Sims & University of Rennes, France & (2, 3) \\
  & ian.sims@univ-rennes1.fr & \\
  Dahbia Talbi & University of Montpellier, France & (2, 3, 5) \\
&  Dahbia.Talbi@univ-montp2.fr & \\
Jonathan Tennyson & University College London, UK & (3, 5, 6)\\
 & j.tennyson@ucl.ac.uk & \\
Roland Wester & University of Innsbruck, Autria & (1, 2, 4, 6) \\
& roland.wester@uibk.ac.at & \\
\enddata
\\
 Field of expertise: (1) Photo and cosmic-ray processes, (2) Neutral-neutral and ion-neutral reactions, (3) Radiative associations, (4) Charge exchange, (5) Dissociative recombination, (6) Associative detachment \\
\end{deluxetable}%

\begin{table*}
\caption{Species added to the kida.uva.2014 network}
\begin{center}
\begin{tabular}{c}
\hline\hline
New species \\
\hline
C$_5$O, C$_6$N, C$_2$H$_6$, C$_7$O, C$_8$N, \\
C$_9$O, C$_{10}$N, HN$_2$O$^+$, HC$_5$O$^+$, C$_6$N$^+$, HC$_6$N$^+$, \\
HC$_7$O$^+$, C$_8$N$^+$, H$_2$C$_6$N$^+$, C$_2$H$_7^+$, HC$_8$N$^+$, HC$_9$O$^+$,\\
 C$_{10}$N$^+$, H$_2$C$_8$N$^+$, HC$_{10}$N$^+$, H$_2$C$_{10}$N$^+$\\
\hline
\end{tabular}
\end{center}
\label{new_spec}
\end{table*}%

\begin{table*}
\caption{Reactions with negative gamma parameters}
\begin{center}
\begin{tabular}{l|l|l}
\hline\hline
Reaction & T range (K) & Reference \\
\hline
 F + H$_2$ $\rightarrow$ H + HF & 10 - 100 & \citet{Neufeld2005} \\
 OH + OH $\rightarrow$ O + H$_2$O & 200 - 350 & datasheet \\
 C + NH$_2$ $\rightarrow$ H + HNC/HNC & 10 - 300 & datasheet \\
 O + OH $\rightarrow$ O$_2$ + H & 150 - 500 & datasheet \\
 H$_2$ + NH$_3^+$ $\rightarrow$ NH$_4^+$ + H & 10 - 20 & datasheet \\
 C + OH $\rightarrow$ H + CO & 10 - 500 & \citet{Zanchet2009}\\
 CH + OCS $\rightarrow$ CO + H + CS & 301 - 667  & \citet{Zabarnick1989}\\
 N + Si $\rightarrow$ SiN + h$\nu$ & 10 - 100 & \citet{Gustafsson2012}�\\
 H + F $\rightarrow$ HF + h$\nu$ & 100 - 2500 & \citet{MagnusGustafsson2014}\\
\hline
\end{tabular}
\end{center}
\label{gamanegative}
\end{table*}%

\begin{table*}
\small
\caption{Species with abundance ratios $\log(\rm X_{new}/ X_{old})$ smaller than -0.3 or larger than 0.3 at $10^5$~yr, and with an abundance in at least one of the models  larger than $10^{-12}$ compared with the total proton density. }
\begin{center}
\begin{tabular}{l|l|l|l|l|l|l|l}
\hline\hline
Species & Ratio & Species & Ratio & Species & Ratio &  Species & Ratio\\
\hline
HC$_2$N$^+$	&	-4.5	&	C$_{10}$H$_2$	&	-1.2	&	N$_2$H$^+$	&	-0.7	&	C$_4$H$^-$	&	-0.4	\\
HC$_8$N	&	-3.1	&	CH$_3$C$_3$N	&	-1.2	&	C$_5$H$_2$	&	-0.7	&	l-C$_3$H	&	-0.4	\\
HC$_6$N	&	-2.9	&	H$_3$C$_4$NH$^+$	&	-1.2	&	C$_5$H$_3$N$^+$	&	-0.7	&	C$_6^-$	&	-0.4	\\
HC$_4$N	&	-2.9	&	C$_8$H$^-$	&	-1.2	&	Mg$^+$	&	0.7	&	Na$^+$	&	0.4	\\
CH$_2$NH	&	-2.4	&	C$_9$	&	-1.1	&	C$_5$H$^-$	&	-0.6	&	HC$_3$NH$^+$	&	-0.4	\\
CCS	&	-2.2	&	C$_8$H	&	-1.1	&	C$_7$	&	-0.6	&	NH$_3$	&	-0.4	\\
HNC	&	-1.7	&	C$_9^-$	&	-1.1	&	C$_7^-$	&	-0.6	&	OCS	&	-0.4	\\
C$_{10}$H	&	-1.7	&	C$_9$N	&	-1.0	&	N$_2$	&	-0.6	&	CCN	&	-0.4	\\
C$_{10}$H$^-$	&	-1.7	&	C$_9$H$_2$N$^+$	&	-1.0	&	OCN 	&	-0.6	&	C$_6$H$_2$	&	-0.4	\\
HC$_9$N	&	-1.6	&	C$_3$H$^+$	&	-1.0	&	C$_5$H	&	-0.6	&	H$_2$CN	&	0.4	\\
NC$_4$N	&	-1.6	&	C$_2$S$^+$	&	-0.9	&	C$_7$H$_2$N$^+$	&	-0.6	&	C$_8$H$_2^+$	&	-0.4	\\
NC$_8$N	&	-1.6	&	C$_7$H	&	-0.9	&	l-C$_3$H$_2$	&	0.6	&	H$_2$CCN	&	-0.4	\\
NC$_6$N	&	-1.6	&	C$_7$H$^-$	&	-0.9	&	CNC$^+$	&	0.6	&	NH$_3^+$	&	-0.4	\\
C$_{11}$	&	-1.6	&	C$_8$H$_2$	&	-0.9	&	C$_3$N$^-$	&	-0.6	&	C$_6$	&	-0.4	\\
CH$_3$C$_7$N	&	-1.5	&	NH	&	-0.9	&	c-C$_3$H	&	-0.6	&	C$_5$N	&	-0.4	\\
HC$_7$N	&	-1.5	&	HC$_2$S$^+$	&	-0.9	&	NH$_2$	&	-0.6	&	HNCCC	&	-0.3	\\
H$_3$C$_7$N$^+$	&	-1.4	&	C$_8^-$	&	-0.9	&	HNO	&	-0.6	&	C$_6$H$_6$	&	-0.3	\\
CH$_3$C$_5$N	&	-1.4	&	C$_7$H$_2^+$	&	-0.8	&	C$_9$H$^+$	&	-0.5	&	C$_5$H$_4^+$	&	-0.3	\\
C$_9$H	&	-1.4	&	C$_{10}$H$^+$	&	-0.8	&	CH$_3$CN$^+$	&	0.5	&	CN$^-$	&	-0.3	\\
HC$_3$N	&	-1.3	&	C$_6$H	&	-0.8	&	SiC$_8$H	&	-0.5	&	HCNCC	&	-0.3	\\
C$_9$H$_2$	&	-1.3	&	CH$_2$CCH	&	-0.8	&	CH$_3$NH$_2$	&	-0.5	&	C$_3$O$^+$	&	0.3	\\
C$_9$H$^-$	&	-1.3	&	C$_6$H$^-$	&	-0.8	&	CH$_3$NH$_3^+$	&	-0.5	&	NS	&	0.3	\\
C$_{10}$	&	-1.3	&	Fe$^+$	&	0.8	&	C$_4$H	&	-0.5	&		&		\\
HC$_5$N	&	-1.3	&	C$_7$N	&	-0.7	&	NH$_4^+$	&	-0.4	&		&		\\
C$_{10}^-$	&	-1.3	&	C$_8$	&	-0.7	&	CH$_3$C$_4$H	&	-0.4	&		&		\\
\hline
\end{tabular}
\end{center}
\label{diff1}
\end{table*}%

\begin{table*}
\caption{Species with abundance ratios $\log(\rm X_{new}/ X_{old})$ smaller than -0.3 or larger than 0.3 at $10^6$~yr, and with an abundance in at least one of the models  larger than $10^{-12}$ compared with the total proton density.}
\begin{center}
\begin{tabular}{l|l|l|l}
\hline\hline
Species & Ratio & Species & Ratio  \\
\hline
l-C$_3$H$_2$  &    1.6 & HC$_3$N   &  0.5 \\
HC$_8$N  &   -1.5 & SiS  &  -0.5 \\
SiC$_8$H  &   -1.4 & C$_3$S   &  0.5 \\
CCS  &   -1.4 & N$_2$O  &   0.5 \\
HC$_6$N   &  -1.2 & CH$_2$CHC$_2$H &   -0.4 \\
NO  &    1.1 & C$_3$O  &   0.4 \\
NC$_8$N   &  -1.1 & SiO$_2$  &  -0.4 \\
SiC$_6$H   &  -1.0 & NO$^+$  &   0.4 \\
H$_2$CCN   &  -1.0 & OCN  &  -0.4 \\
NC$_6$N &   -0.9 & C$_4$H  &  -0.4 \\
C$_6$H$_6$  &  -0.8 & H$_2$NC$^+$   &  0.3 \\
C$_3$N   &  0.8 & l-C$_3$H$_3^+$  &   0.3 \\
OCS   & -0.7 & NH$_2$ &   -0.3 \\
CNC$^+$   &  0.7 & HNO  & -0.3 \\
HNCCC  &   0.7 & C$_3$  &   0.3 \\
HCCNC  &   0.6 & N  &   0.3 \\
HC$_4$N  &  -0.6 & l-C$_3$H &   -0.3 \\
c-C$_3$H  &  -0.6 & C$_4$H$_2$  &  -0.3 \\
\hline
\end{tabular}
\end{center}
\label{diff2}
\end{table*}%


\end{document}